\documentclass[aps,prr,twocolumn,amsmath,amssymb,superscriptaddress,longbibliography]{revtex4-1}
\usepackage[english]{babel}
\usepackage{xcolor}
\usepackage{amssymb}
\usepackage{dcolumn}
\usepackage{bm}
\usepackage{graphicx}
\usepackage{amsmath}
\usepackage{braket}

\usepackage{graphicx}        

\graphicspath{{pict/}{}}

\usepackage[normalem]{ulem}

\usepackage{dcolumn}
\usepackage{bm}
\usepackage[pdfstartview=FitH, CJKbookmarks=true, bookmarksnumbered=true, bookmarksopen=true, colorlinks=true, pdfborder=001, citecolor=blue, linkcolor=blue, urlcolor=blue, linktocpage=true] {hyperref}

\begin{document}


\title{Deterministic multiphoton bundle emission via interference-interaction control}

\author{Jing Tang}
\email{jingtang@gdut.edu.cn}
\affiliation{School of Physics and Optoelectronic Engineering, Guangdong University of Technology, Guangzhou 510006, China}
\affiliation{Guangdong Provincial Key Laboratory of Sensing Physics and System Integration Applications, Guangdong University of Technology, Guangzhou, 510006, China}

\author{Yuangang Deng}
\email{dengyg3@mail.sysu.edu.cn}
\affiliation{Guangdong Provincial Key Laboratory of Quantum Metrology and Sensing $\&$ School of Physics and Astronomy, Sun Yat-Sen University (Zhuhai Campus), Zhuhai 519082, China}

\date{\today}

\begin{abstract}
The controlled generation of nonclassical light beyond single photons remains a central challenge in quantum optics, due to the difficulty of enhancing multiphoton processes while suppressing lower-order excitations. Here we propose an interference-interaction-engineered scheme for programmable few-photon emission in a cavity-QED system of three atoms coupled to orthogonal cavity modes. By adiabatically eliminating an auxiliary Fabry-P\'erot cavity, we generate a tunable cavity-mediated spin-exchange interaction $\chi$, which, combined with a controllable geometric phase $\phi$, reshapes the many-body dressed-state spectrum. This interplay enables selective addressing of excitation manifolds ($N=1,2,3$), establishing a direct mapping between excitation structure and photon-emission channels. For $\phi=0$, constructive interference enhances the spectral anharmonicity of low-excitation manifolds, yielding tunable single- and two-photon emission associated with the $N=1$ and $N=2$ manifolds. In contrast, for $\phi=2\pi/3$, destructive interference suppresses lower-order excitation pathways and activates a resonant three-photon channel originating from the $N=3$ manifold. Importantly, the cavity-mediated interaction $\chi$ further enhances spectral separation between manifolds, enabling a substantial improvement in multiphoton purity while maintaining a sizable photon population. We demonstrate a three-order-of-magnitude enhancement in two-photon purity and more than two orders of magnitude improvement in three-photon emission. Our results establish a unified interference-interaction framework in which effective optical nonlinearities can be programmably engineered through phase and interaction, providing a scalable route toward high-purity multiphoton sources and programmable quantum photonic devices. 
\end{abstract}

\maketitle
\section{Introduction}
Nonclassical light sources, ranging from antibunched single photons to strongly correlated multiphoton states, constitute key resources for quantum communication, computation, and precision metrology~\cite{Duan01, Kimble08, Giovannetti2011, Aspuru-Guzik2012}. In cavity quantum electrodynamics (cQED), such states are typically realized via photon blockade, where optical nonlinearities generate anharmonic energy spectra that suppress unwanted multiphoton excitations~\cite{Imamoglu97, Hennessy07, PhysRevX.5.031028}. While single-photon blockade has been extensively demonstrated~\cite{Birnbaum2005, Faraon08, Fink08, Reinhard2012, Kai15, tang21}, extending this mechanism to controllable and high-purity multiphoton emission, such as photon pairs and higher-order photon bundles, remains a major challenge due to the difficulty of simultaneously enhancing higher-order processes while suppressing lower-order excitations.

Existing approaches to multiphoton generation rely either on strong intrinsic nonlinearities or on frequency-selective multiphoton resonances~\cite{PhysRevLett.122.123604, Muller2014, munoz2014emitters, PhysRevLett.118.133604, PhysRevLett.127.073602}. However, these schemes typically require stringent conditions, such as strong coupling or large detuning, which limit their tunability and scalability~\cite{xu18, zhao2025tunable, PhysRevResearch.6.033247, liu2023deterministic, PhysRevLett.133.043601}.  An alternative route is offered by interference-based mechanisms, where quantum pathways are engineered to selectively suppress or enhance specific excitation processes~\cite{Bamba2011, Tang15, Snijders18, Vaneph18, zhou2025universal}. Despite their flexibility, most existing proposals rely on multi-tone driving, Raman transitions, or complex level structures, which complicate experimental implementation~\cite{Liew10, Majumdar2012, Flayac17, Tang19, PhysRevLett.127.240402, chen2022photon, tang2022strong}. In parallel, photon-mediated long-range interactions, typically realized via dissipative or auxiliary cavity modes, have emerged as a powerful tool to reshape excitation spectra and induce collective many-body dynamics in multi-emitter cQED systems~\cite{Hartmann06, van2013photon, Chang14, aron2016photon, vaidya2018tunable, RevModPhys.87.1379, ren2023enhancing}. These interactions can enhance spectral anharmonicity and enable collective control of light-matter coupling. However, a unified framework that combines interference and interaction to enable programmable and selective multiphoton emission remains largely unexplored~\cite{PhysRevA.110.030101, chakram2022multimode, carl2023phases}.

In this work, we develop a unified interference-interaction framework for programmable few-photon emission in a cavity-QED system of three two-level atoms coupled to orthogonal cavity modes. By operating one cavity in a far-detuned dispersive regime and adiabatically eliminating it, we engineer a tunable cavity-mediated spin-exchange interaction (SEI) $\chi$, arising from virtual photon exchange processes mediated by the auxiliary cavity mode. Combined with a controllable geometric phase $\phi$ determined by the atomic spatial configuration, this setup enables simultaneous control of quantum interference and many-body excitation spectra. The interplay between phase-controlled interference and interaction-induced spectral anharmonicity enables selective addressing of distinct excitation manifolds, providing a direct mapping between excitation structure and photon-emission channels. With this framework, photon emission is governed by manifold-resolved resonances: constructive interference enhances the effective anharmonicity of low-excitation manifolds, while destructive interference suppresses competing excitation pathways and redistributes spectral weight toward higher-order manifolds. As a result, distinct photon-emission channels can be selectively activated by tuning phase $\phi$ and the SEI strength $\chi$. Physically, this mechanism establishes a direct mapping between excitation manifolds and photon-emission processes. Under constructive interference, the low-excitation manifolds ($N=1,2$) dominate, giving rise to single- and two-photon emission, whereas suppression of lower-order pathways isolates the higher-excitation manifold ($N=3$), enabling three-photon emission. 

The cavity-mediated SEI further enhances spectral separation between different excitation manifolds, thereby enhancing emission selectivity and effectively amplifying optical nonlinearity without relying on intrinsic material anharmonicity. As a result, both two- and three-photon bundle emissions exhibit orders-of-magnitude improvements in purity while maintaining a sizable photon population. These results establish a unified interference-interaction paradigm in which effective optical nonlinearities can be programmably engineered through phase control and photon-mediated interactions, providing a general route toward tunable nonlinear optical responses beyond intrinsic material constraints. The scheme is compact, requires only resonant driving and geometric phase control, and is compatible with existing cavity-QED platforms, including neutral atoms, Rydberg systems, and alkaline-earth implementations~\cite{tiarks2019photon, norcia2018cavity, lu2025chiral}. Moreover, it naturally extends to larger emitter arrays and multimode cavity architectures, providing a scalable route toward programmable quantum photonic devices and controlled many-body light generation~\cite{wang2025scalable, tang2024tunable}. 

\begin{figure}[ht]
\includegraphics[width=1\columnwidth]{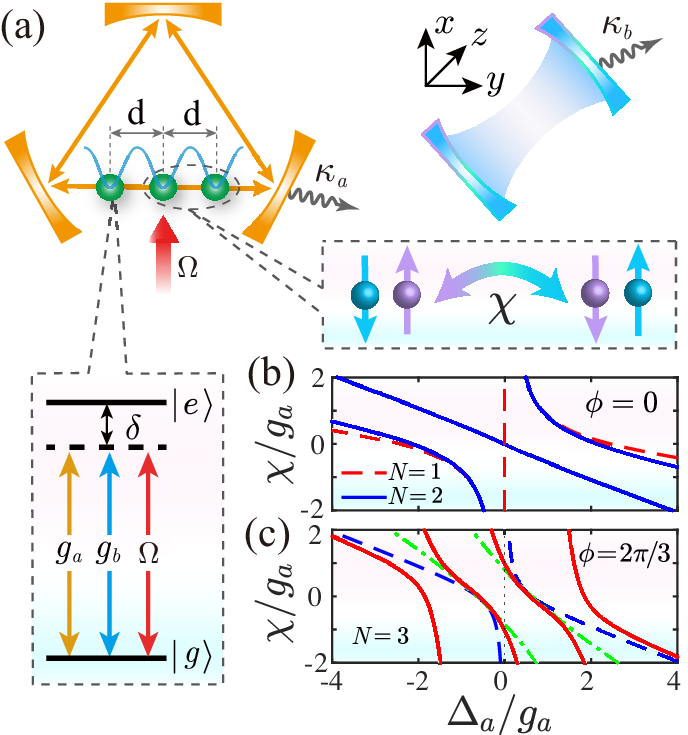}
\caption{(a) Schematic of cavity-coupled three atom array with SEI.  The dashed boxes illustrate the single-atom level structure and tunable long-range SEI induced by the far-detuned auxiliary Fabry-P\'erot cavity. The relative atomic positions imprint a controllable phase $\phi$, enabling interference-interaction engineering of photon-emission processes. (b) Manifold-resolved zero-energy resonance curves in the $(\chi,\Delta_a)$ plane at $\phi=0$, obtained from $N=1$ and $N=2$ excitation manifolds, identifying the resonance conditions for single- and two-photon emission. (c) Zero-energy resonance curves at $\phi=2\pi/3$, derived from the $N=3$ manifold. Destructive interference suppresses lower-order excitation pathways and isolates the higher-excitation manifold, enabling selective activation of three-photon resonance. The detuning is fixed at $\delta/\Delta_a=0.5$.}
\label{model1}
\end{figure}

\section{Model and energy spectrum}
We consider three identical two-level atoms (ground state $|g\rangle$ and excited state $|e\rangle$) coupled to two single-mode optical cavities: a ring cavity and a Fabry-P\'erot cavity with orthogonal polarizations, as illustrated in Fig.~\ref{model1}(a). This cross-cavity configuration establishes a phase-controllable interference channel between atomic transitions, enabling coherent control of photon-emission pathways.  The cavity modes $\hat{a}$ and $\hat{b}$ have resonance frequency $\omega_c$ and decay rates $\kappa_a$ and $\kappa_b$, respectively. The atoms are separated by a distance $d$, which induces a controllable relative phase $\phi = k_l d$, where $k_l$ is the wavevector associated with cavity mode $\hat{a}$. A resonant laser with Rabi frequency $\Omega$ drives the $|g\rangle \leftrightarrow |e\rangle$ transition, forming a compact and highly tunable platform for phase-engineered quantum interference and photon statistics control.

In the rotating frame, the system Hamiltonian is given by
\begin{align}
{\cal \hat{H}'} 
&= \Delta_a \hat{a}^\dagger \hat{a} + \Delta_b \hat{b}^\dagger \hat{b} 
+ \frac{1}{2} \sum_{j=1}^{3} (\delta \sigma_j^z + \Omega \sigma_j^x) \nonumber \\
&\quad + g_a \hat{a}^\dagger (e^{i\phi}\sigma_1^- + \sigma_2^- + e^{-i\phi}\sigma_3^-) \nonumber \\
&\quad + g_b \hat{b}^\dagger (\sigma_1^- + \sigma_2^- + \sigma_3^-) + {\rm H.c.}
\label{hami1}
\end{align}
where $\hat{a}$ and $\hat{b}$ denote annihilation operators of the two cavity modes with the tunable light-cavity detuning $\Delta_a$ and $\Delta_b$,  $\sigma^{x,y,z}$  are Pauli operators for atom $j$, and $\delta$ is the single-photon detuning. $g_a$ and $g_b$ quantify the single-photon atom-cavity coupling strengths, while the phase factor $\phi$ encodes the relative atomic positions along mode $\hat{a}$.

In the large dispersive regime of mode $|\Delta_b| \gg g_b, \kappa_b$, the auxiliary cavity can be adiabatically eliminated. The system is then described by the effective Hamiltonian
\begin{align}
{\cal \hat{H}} 
&= \Delta_a \hat{a}^\dagger \hat{a} 
+ \frac{1}{2} \sum_{j=1}^{3} (\delta \sigma_j^z + \Omega \sigma_j^x)
+ \chi \sum_{j,k} \sigma_j^+ \sigma_k^- \nonumber \\
&\quad + g_a \hat{a}^\dagger (e^{i\phi}\sigma_1^- + \sigma_2^- + e^{-i\phi}\sigma_3^-) 
+ {\rm H.c.},
\label{Hami2}
\end{align}
where $\chi = -g_b^2 \Delta_b / (\Delta_b^2 + \kappa_b^2)$ denotes a tunable cavity-mediated SEI arising from virtual photon exchange via mode $\hat{b}$. The SEI induces collective, interaction-dependent energy shifts that lift the degeneracy between different excitation manifolds, thereby generating an effective many-body anharmonic ladder. This reduction maps the original cross-cavity configuration onto an effective single-mode cavity QED system with engineered long-range interaction and phase-dependent interference, providing a minimal platform for programmable many-body light-matter coupling.

To elucidate the mechanism of selective multiphoton emission, we analyze the excitation spectrum of the effective Hamiltonian in Eq.~(\ref{Hami2}). The cavity-mediated SEI $\chi$ induces spectral anharmonicity, while the phase $\phi$ controls interference between different excitation pathways. The interplay between these two ingredients reshapes the structure of excitation manifolds.

Neglecting the weak coherent drive, the total excitation number
\begin{equation}
\hat N=\hat a^\dagger\hat a+\sum_{j=1}^3\sigma_j^+\sigma_j^-
\end{equation}
is conserved, i.e., $[\hat{\cal H},\hat N]=0$. The dynamics can therefore be decomposed into subspaces with fixed excitation number $N$. In each manifold, the Hamiltonian can be expressed in the ordered basis
\begin{equation}
\begin{aligned}
\Psi = [&
|n\!-\!1,g,g,e\rangle,\;
|n\!-\!1,g,e,g\rangle,\;
|n\!-\!1,e,g,g\rangle,\\
&|n\!-\!2,e,e,g\rangle,\;
|n\!-\!2,e,g,e\rangle,\;
|n\!-\!2,g,e,e\rangle,\\
&|n\!-\!3,e,e,e\rangle,\;
|n,g,g,g\rangle
]^T ,
\end{aligned}
\end{equation}
leading to the matrix representation $\mathcal{M}$
\begin{widetext}
\begin{equation}
\begin{pmatrix}
E_1 & \chi & \chi & 0 & g_a\sqrt{n-1} & g_a\sqrt{n-1}e^{-i\phi} & 0 & g_a\sqrt{n}e^{i\phi} \\
\chi & E_1 & \chi & g_a\sqrt{n-1}e^{i\phi} & 0 & g_a\sqrt{n-1}e^{-i\phi} & 0 & g_a\sqrt{n} \\
\chi & \chi & E_1 & g_a\sqrt{n-1}e^{i\phi} & g_a\sqrt{n-1} & 0 & 0 & g_a\sqrt{n}e^{-i\phi} \\
0 & g_a\sqrt{n-1}e^{-i\phi} & g_a\sqrt{n-1}e^{-i\phi} & E_2 & \chi & \chi & g_a\sqrt{n-2}e^{-i\phi} & 0 \\
g_a\sqrt{n-1} & 0 & g_a\sqrt{n-1} & \chi & E_2 & \chi & g_a\sqrt{n-2} & 0 \\
g_a\sqrt{n-1}e^{i\phi} & g_a\sqrt{n-1}e^{i\phi} & 0 & \chi & \chi & E_2 & g_a\sqrt{n-2}e^{i\phi} & 0 \\
0 & 0 & 0 & g_a\sqrt{n-2}e^{i\phi} & g_a\sqrt{n-2} & g_a\sqrt{n-2}e^{-i\phi} & E_3 & 0 \\
g_a\sqrt{n}e^{-i\phi} & g_a\sqrt{n} & g_a\sqrt{n}e^{i\phi} & 0 & 0 & 0 & 0 & E_4
\end{pmatrix}
\label{mat_M_corrected}
\end{equation}
\end{widetext}
where
\begin{align}
E_1 &= (n-1)\Delta_a + \delta + \chi, \nonumber \\
E_2 &= (n-2)\Delta_a + 2(\delta + \chi),\nonumber \\
E_3 &= (n-3)\Delta_a + 3(\delta + \chi),\nonumber \\
E_4 &= n\Delta_a,
\end{align}
which correspond to the single-, double-, triple-, and zero-atomic-excitation manifolds, respectively.

This block-structured Hamiltonian reveals how SEI-induced anharmonicity and phase-controlled interference jointly determine the excitation spectrum. The interaction $\chi$ lifts the degeneracy between different excitation manifolds, while the phase $\phi$ redistributes coupling strengths between excitation pathways. As a result, the excitation manifolds become spectrally distinguishable and can be selectively addressed.

To identify the corresponding resonance conditions, we analyze the zero-energy solutions of each excitation manifold by imposing $\det(\mathcal{M}_N)=0$ (see Appendix for details). These conditions define manifold-resolved resonances that determine the dominant excitation pathways.

We first consider the representative parameter regime $\phi=0$ and $\delta=\Delta_a/2$, where the system exhibits full permutation symmetry. In the single-excitation manifold ($N=1$),  the spectrum shows an interaction-modified vacuum Rabi splitting with resonance condition
\begin{equation}
\Delta_a=-3\chi\pm\sqrt{9\chi^2+6g_a^2}.
\end{equation}
This interaction-induced shift enhances spectral anharmonicity and establishes a well-isolated resonance associated with single-photon excitation.

In the double-excitation manifold ($N=2$), the hybridization between atomic and photonic excitations gives rise to a cubic resonance condition
\begin{equation}
3\Delta_a^3
+18\chi \Delta_a^2
+\left(24\chi^2-14g_a^2\right)\Delta_a
-24g_a^2\chi
=0 ,
\end{equation}
which reflects the increased complexity of multi-excitation dressed states. The corresponding resonance branches define the conditions for two-photon excitation.

In contrast, for $\phi=2\pi/3$, the phase factors satisfy $1+e^{i\phi}+e^{-i\phi}=0$, resulting in complete destructive interference of the collective coupling in the single-excitation channel. Consequently, the $N=1$ and $N=2$ excitation pathways are suppressed, and the system preferentially accesses higher-excitation manifolds. Therefore the dominant resonances emerge in the $N=3$ manifold, governed by
\begin{align}
5\Delta_a^2+10\chi\Delta_a-2g_a^2 &= 0,\nonumber \\
10\Delta_a^2+17\chi\Delta_a+6\chi^2-4g_a^2 &= 0, \nonumber \\
10\Delta_a^4+25\chi\Delta_a^3+10\chi^2\Delta_a^2
-38g_a^2\Delta_a^2
-62\chi g_a^2\Delta_a
\nonumber\\
\qquad
-12\chi^2 g_a^2
+12g_a^4 &= 0,
\end{align}
These branches define the resonance structure of the $N=3$ manifold and enable selective activation of the three-photon excitation channel.

Overall, the excitation spectrum exhibits a clear manifold-resolved structure: for $\phi=0$, constructive interference enhances the spectral separation of the $N=1$ and $N=2$ manifolds, supporting single- and two-photon processes [Fig.~\ref{model1}(b)]; whereas for $\phi=2\pi/3$, destructive interference suppresses these lower-order channels and promotes higher-order resonances in the $N=3$ manifold[Fig.~\ref{model1}(c)]. This establishes a direct mapping between excitation manifolds and photon-emission channels.
Therefore, the interplay between phase-controlled interference and cavity-mediated SEI provides a unified mechanism for selectively addressing different excitation manifolds and realizing programmable few-photon emission, as confirmed by the photon statistics in the following.

To characterize the quantum correlations of the emitted light, we describe the system dynamics using a Lindblad master equation,
\begin{align}
\frac{d\rho}{dt}
= -\frac{i}{\hbar}[\hat{\cal H},\rho]
+ \kappa_a \mathcal{D}[\hat a]\rho
+ (\gamma+\gamma_e)\sum_{j=1}^{3} \mathcal{D}[\hat\sigma_j^-]\rho,
\end{align}
where $\rho$ is the density matrix and $\mathcal{D}[\hat o]\rho=\hat o\rho \hat o^\dagger-(\hat o^\dagger\hat o\rho+\rho\hat o^\dagger\hat o)/2$ denotes the standard Lindblad dissipator.

The effective decay rate 
\begin{equation}
\gamma_e=\frac{\kappa_b g_b^2}{\Delta_b^2+\kappa_b^2}
\end{equation}
arises from the adiabatic elimination of the auxiliary cavity mode, which mediates collective atomic interactions. The steady-state properties of the system are obtained by numerically solving the master equation under the condition $d\hat\rho/dt=0$. From the steady-state density matrix $\hat\rho_s$, we extract both the intracavity photon number $n_s = \mathrm{Tr}(\hat a^\dagger \hat a \hat\rho_s)$ and the photon correlation functions.

The nature of the emitted quantum light is characterized by the normalized higher-order correlation functions~\cite{munoz2014emitters,del2012theory},
\begin{align}
g_n^{(k)}(\tau_1,\dots,\tau_k)
= \frac{\left\langle \prod_{i=1}^k \left[\hat a^\dagger(\tau_i)\right]^n \prod_{i=1}^k \left[\hat a(\tau_i)\right]^n \right\rangle}
{\prod_{i=1}^k \left\langle \left[\hat a^\dagger(\tau_i)\right]^n \left[\hat a(\tau_i)\right]^n \right\rangle},
\end{align}
with $\tau_1 \leq \dots \leq \tau_k$. 

Single-photon blockade is identified by $g_1^{(2)}(0) < 1$ together with antibunching behavior $g_1^{(2)}(0) < g_1^{(2)}(\tau)$. For higher-order processes, $n$-photon blockade ($n\ge2$) requires $g_1^{(n)}(0) > 1$ and $g_1^{(n+1)}(0) < 1$~\cite{PhysRevLett.118.133604,PhysRevLett.121.153601}. In the regime of photon-bundle emission, additional temporal correlations are required: $g_1^{(2)}(0) > g_1^{(2)}(\tau)$ ensures bunching within each bundle, while $g_n^{(2)}(0) < g_n^{(2)}(\tau)$ indicates antibunching between successive bundles~\cite{munoz2014emitters,Chang16}.

The proposed scheme can be implemented  in cavity-QED platforms based on Rydberg atoms~\cite{PhysRevLett.104.195302, madjarov2020high} or alkaline-earth-metal atoms~\cite{PhysRevLett.117.220401, kolkowitz2017spin, PhysRevLett.118.263601, bromley2018dynamics} coupled to optical cavities~\cite{kroeze2023high, karg2020light}, leveraging their favorable level structures and long coherence times. In particular, the use of a low-finesse cavity with a large decay rate $\kappa_b$ enables the realization of strong effective interactions, which are essential for accessing the nonlinear photon-emission regimes explored in this work. Recent demonstrations of on-chip generation and tomography of nonclassical light further support the feasibility of implementing phase-programmable quantum light sources in integrated photonic platforms~\cite{park2024single}. 

For numerical simulations, the cavity decay rate is set to $\kappa=2\pi \times 150$ kHz, serving as the energy unit,  which capability has been demonstrated in current atom-cavity experiments~\cite{Leonard2017}. The spontaneous decay rate of the long-lived excited state is fixed at $\gamma/\kappa=0.2$~\cite{PhysRevLett.122.123604, Peng22}. We adopt a moderate atom-cavity coupling strength $g/\kappa=10$ and a weak atomic drive of $\Omega/\kappa=0.5$ are adopted, which are  well within current experimental capabilities~\cite{PhysRevLett.118.133604}.

\section{Results and Discussion}
\subsection{Phase-controlled single- to three-photon emission}
\begin{figure}[ht]
\includegraphics[width=0.98\columnwidth]{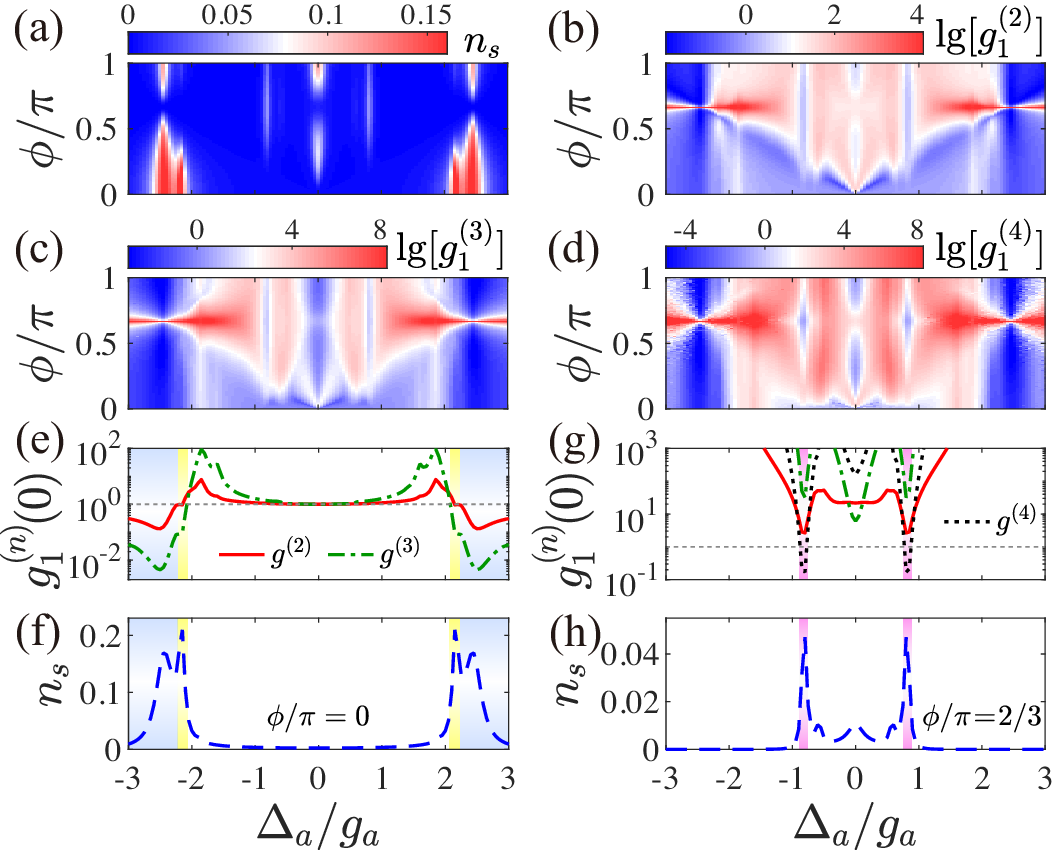}
\caption{Phase-controlled photon statistics in the absence of interaction ($\chi/g_a=0$). (a) Steady-state photon number $n_s$, and (b)-(d) equal-time correlation functions lg$[g_1^{(2)}(0)]$, lg$[g_1^{(3)}(0)]$, lg$[g_1^{(4)}(0)]$ as functions of the cavity detuning $\Delta_a$. 
(e,f) and (g,h) show the corresponding photon correlations [$g_1^{(n)}(0), n=2,3,4$] and photon number $n_s$ for $\phi/\pi=0$ and $2/3$, respectively. 
For $\phi/\pi=0$, constructive interference enhances the spectral separation of the $N=1$ and $N=2$ manifolds, giving rise to single-photon blockade (light-blue regions) and two-photon blockade (light-yellow regions). In contrast, for $\phi/\pi=2/3$, only three-photon blockade 
(pink regions) appears due to phase-controlled interference. The detuning is fixed at $\delta/\Delta_a=0.5$.}
\label{phi}
\end{figure}

The impact of manifold-resolved resonances on photon statistics is illustrated in Fig.~\ref{phi}, where we first consider the noninteracting case ($\chi=0$). Figures~\ref{phi}(a-d) show the steady-state photon number $n_s$ and the equal-time correlation functions $g_1^{(2)}(0)$, $g_1^{(3)}(0)$, and $g_1^{(4)}(0)$ (in logarithmic scale) as functions of the detuning $\Delta_a$ and the phase $\phi$. The spectra exhibit a clear symmetry with respect to $\Delta_a=0$, reflecting the underlying symmetry of the excitation structure.

For $\phi=0$, the emission is dominated by the $N=1$ and $N=2$ manifolds. As shown in Figs.~\ref{phi}(e) and \ref{phi}(f), around the single-excitation resonance $\Delta_a/g_a=\pm2.5$, strong antibunching with $g_1^{(2)}(0)=1.33\times10^{-1}$ is observed, accompanied by a finite photon population, indicating high-purity single-photon emission (light-blue regions). This behavior originates from the anharmonic vacuum Rabi splitting in the $N=1$ manifold, which energetically isolates the single-photon transition and suppresses higher-order excitations. At the two-excitation resonance $\Delta_a/g_a=\pm2.15$, the system exhibits enhanced two-photon correlations, with $g_1^{(2)}(0)\approx1$, $g_1^{(3)}(0)=9.4\times10^{-2}$, and $n_s=0.21$, signaling the activation of the two-photon emission channel (light-yellow regions). Here, the spectral proximity of the $N=2$ manifold enables sequential two-photon excitation, while residual anharmonicity partially suppresses higher-order processes.

\begin{figure}[ht]
\includegraphics[width=1\columnwidth]{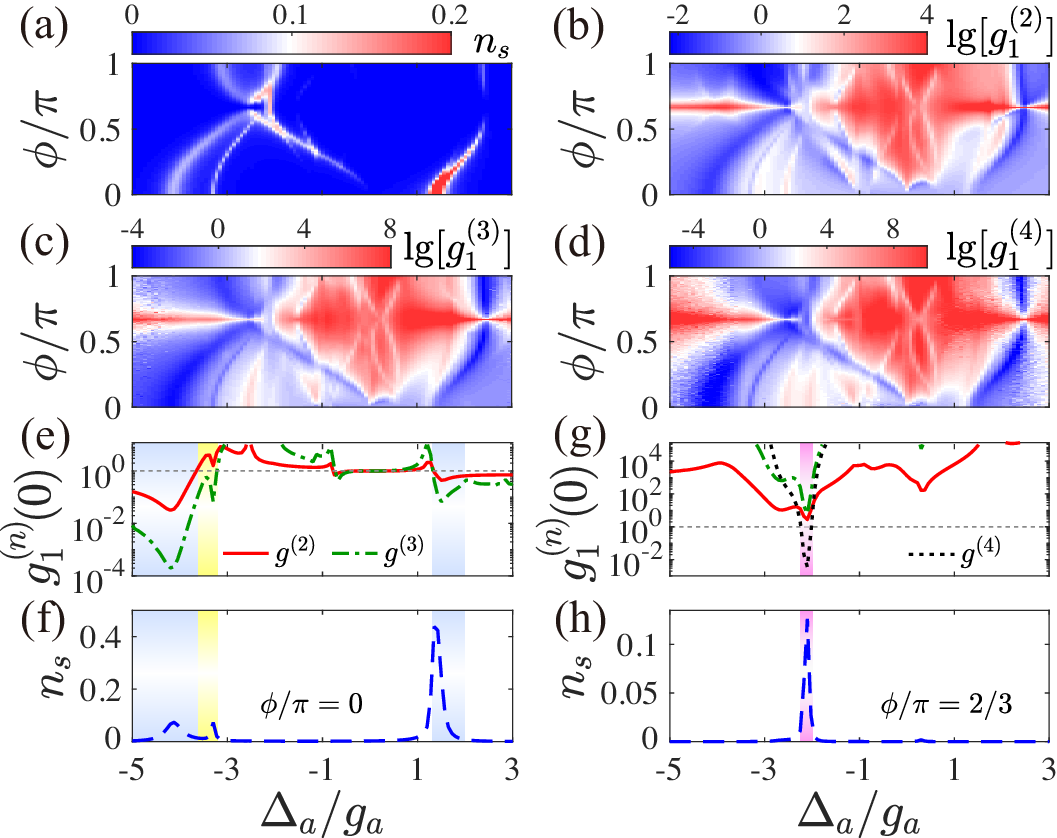}
\caption{Phase-space mapping of photon correlations and emission statistics in the presence of interatomic interactions ($\chi/g_a=0.45$). 
(a) Steady-state photon number $n_s$, and (b)-(d) equal-time correlation functions lg$[g_1^{(2)}(0)]$, lg$[g_1^{(3)}(0)]$, lg$[g_1^{(4)}(0)]$ as functions of the cavity detuning $\Delta_a$. (e,f) and (g,h) display the corresponding photon correlations $g_1^{(n)}(0)$ and photon number $n_s$ 
for $\phi/\pi=0$ and $2/3$, respectively. In contrast to noninteracting case, the spin-exchange interaction $\chi$ significantly enhances photon blockade and strengthens multiphoton correlations. This behavior highlights the cooperative interplay between phase-controlled interference and interaction-induced spectral reshaping in engineering high-order nonclassical light. Other parameter is $\delta/\Delta_a=0.5$.}
\label{phi_V}
\end{figure}

In contrast, for $\phi=2\pi/3$, destructive interference suppresses the coupling to the lower-excitation manifolds ($N=1,2$), effectively blocking the corresponding excitation pathways. Consequently, the spectral weight is redistributed toward higher-excitation manifolds, and a dominant resonance emerges in the $N=3$ manifold at $\Delta_a/g_a=\pm0.8$ (pink regions). As shown in Figs.~\ref{phi}(g) and \ref{phi}(h), this regime is characterized by strong three-photon correlations, with $g_1^{(2)}(0)=4.23$, $g_1^{(3)}(0)=31.93$, $g_1^{(4)}(0)=1.7\times10^{-1}$, and $n_s=4.7\times10^{-2}$. This behavior originates from interference-induced pathway selection: the phase $\phi=2\pi/3$ enforces destructive interference among single- and two-excitation channels while leaving higher-order collective excitations accessible. As a result, lower-order processes are effectively filtered out and the system selectively favors three-photon excitation. These results demonstrate that phase-controlled interference alone can redistribute excitation pathways across different manifolds, enabling selective activation of distinct photon-emission channels.

We now include the cavity-mediated spin-exchange interaction and set $\chi/g_a=0.45$, as shown in Fig.~\ref{phi_V}. In contrast to the symmetric response of the noninteracting case, the photon statistics become strongly asymmetric with respect to the detuning $\Delta_a$. This asymmetry arises from interaction-induced energy shifts, which break the inherent $\Delta_a\leftrightarrow -\Delta_a$ symmetry of the excitation spectrum and modify the resonance conditions.

\begin{figure*}[t!]  
\includegraphics[width=2\columnwidth]{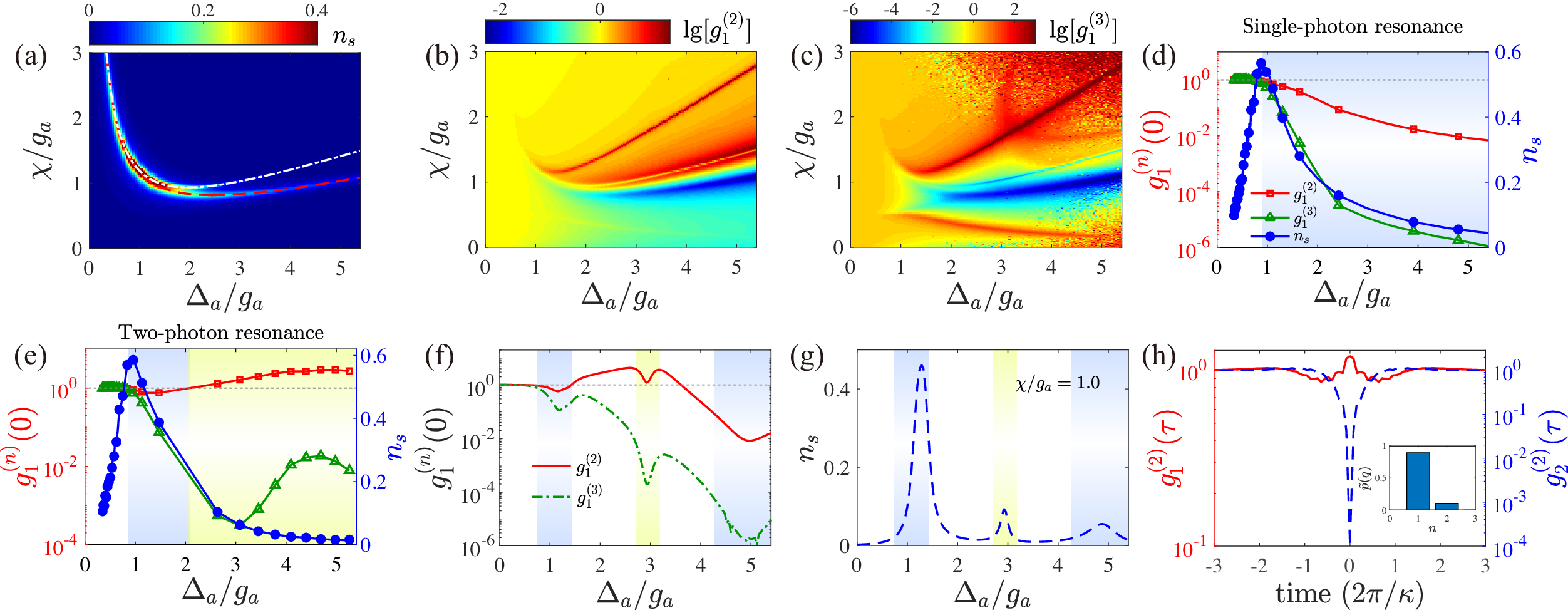}
\caption{Phase diagram of photon correlations and emission statistics in the $\chi$-$\Delta_a$ parameter plane for $\phi=0$. 
(a) Steady-state photon number $n_s$, and the correlation functions lg$[g_1^{(2)}(0)]$ (b), lg$[g_1^{(3)}(0)]$ (c) as functions of the cavity detuning $\Delta_a$ and interaction strength $\chi$, with fixed $\delta/\Delta_c=-0.5$. The red and white curves  in (a) denote the analytical resonance conditions of the energy spectrum for the single-photon and two-photon resonances, respectively. (d-e) Photon correlations $g_1^{(2)}(0)$, $g_1^{(3)}(0)$, and photon number $n_s$ extracted along the two resonance branches as functions of $\Delta_a$, highlighting the single-photon and two-photon emission regimes. 
(f,g) Photon correlation functions $g_1^{(2)}(0)$ and $g_1^{(3)}(0)$ together with the steady-state photon number $n_s$ as functions of $\Delta_a$ at $\chi/g_a=1.0$. Pronounced photon antibunching occurs at $\Delta_a/g_a=1.28$ and $4.88$, corresponding to single-photon emission, while a strong two-photon resonance appears at $\Delta_a/g_a=2.92$. 
(h) Time-dependent second-order correlation functions $g_1^{(2)}(\tau)$ (red line) and $g_2^{(2)}(\tau)$ (blue line) at the two-photon resonance point $\Delta_a/g_a=2.92$. 
The inset shows the steady-state photon-number distribution $\tilde{p}(q)$ at the same point, demonstrating dominant two-photon occupation and suppression of single-photon processes.}
\label{V_Deltac}
\end{figure*}

For $\phi=0$, the single- and two-photon emission regimes associated with the $N=1$ and $N=2$ manifolds persist, but exhibit markedly improved purity. As can be seen in Figs.~\ref{phi_V}(e) and ~\ref{phi_V}(f), the antibunching at the single-photon resonance ($\Delta_a/g_a=-4.2$) is significantly enhanced, yielding $g_1^{(2)}(0)=3.2\times10^{-2}$, while the two-photon emission region becomes more clearly resolved, with $g_1^{(2)}(0)=1.6$ and $g_1^{(3)}(0)=6.8\times10^{-2}$. This improvement originates from interaction-induced spectral anharmonicity, which increases the separation between adjacent excitation manifolds and suppresses unwanted higher-order transitions. As a result, excitation pathways become more selective, leading to higher emission purity. Notably, this enhanced selectivity is accompanied by a reduction in the photon number, reflecting a trade-off between emission efficiency and spectral filtering.

The most pronounced effect occurs in the $\phi=2\pi/3$ regime. As shown in Figs.~\ref{phi_V}(g) and ~\ref{phi_V}(h), the three-photon emission associated with the $N=3$ manifold is significantly enhanced, with $g_1^{(2)}(0)=2.77$, $g_1^{(3)}(0)=6.19$, $g_1^{(4)}(0)=3.5\times10^{-3}$, and $n_s=1.26\times10^{-1}$. Compared to the noninteracting case, the photon number is substantially increased while higher-order correlations are strongly suppressed, indicating a simultaneous improvement in both efficiency and purity.

Physically, the role of the SEI $\chi$ is to reinforce the spectral isolation between different excitation manifolds. While phase-controlled interference determines which excitation pathways are allowed or suppressed, the interaction enhances the effective nonlinearity by increasing the separation between manifolds. This reduces spectral overlap and suppresses leakage into undesired excitation channels. As a result, a cooperative mechanism emerges: interference selects the target excitation manifold, while the SEI stabilizes and amplifies the corresponding photon-emission process. This synergy enables robust and high-purity multiphoton generation. Importantly, this mechanism does not rely on intrinsically strong nonlinearities, but instead exploits interaction-engineered spectral structure for programmable quantum light generation.

\subsection{Interaction-enhanced two-photon bundle emission}
We next identify the optimal regime for two-photon bundle emission by tuning the system parameters along the two-photon resonance, focusing on the case $\phi=0$. By scanning the detuning $\delta$, we find that the strongest two-photon response occurs near $\delta=-\Delta_a/2$, where the $N=2$ manifold becomes resonant while the $N=1$ channel is effectively suppressed.

Figures~\ref{V_Deltac}(a-c) show the steady-state photon number $n_s$ and the correlation functions $g_1^{(2)}(0)$ and $g_1^{(3)}(0)$ (in logarithmic scale) as functions of the detuning $\Delta_a$ and the interaction strength $\chi$. Two distinct resonance branches are clearly visible: the single-photon resonance (red curve), characterized by strong antibunching with $g_1^{(2)}(0)<1$, and the two-photon resonance (white curve), where $g_1^{(2)}(0)>1$ and $g_1^{(3)}(0)<1$, indicating the activation of the two-photon emission channel. The excellent agreement between the analytical resonance conditions (see Appendix for details) and the numerical results confirms the validity of the manifold-resolved description.

\begin{figure*}[ht]
\includegraphics[width=2\columnwidth]{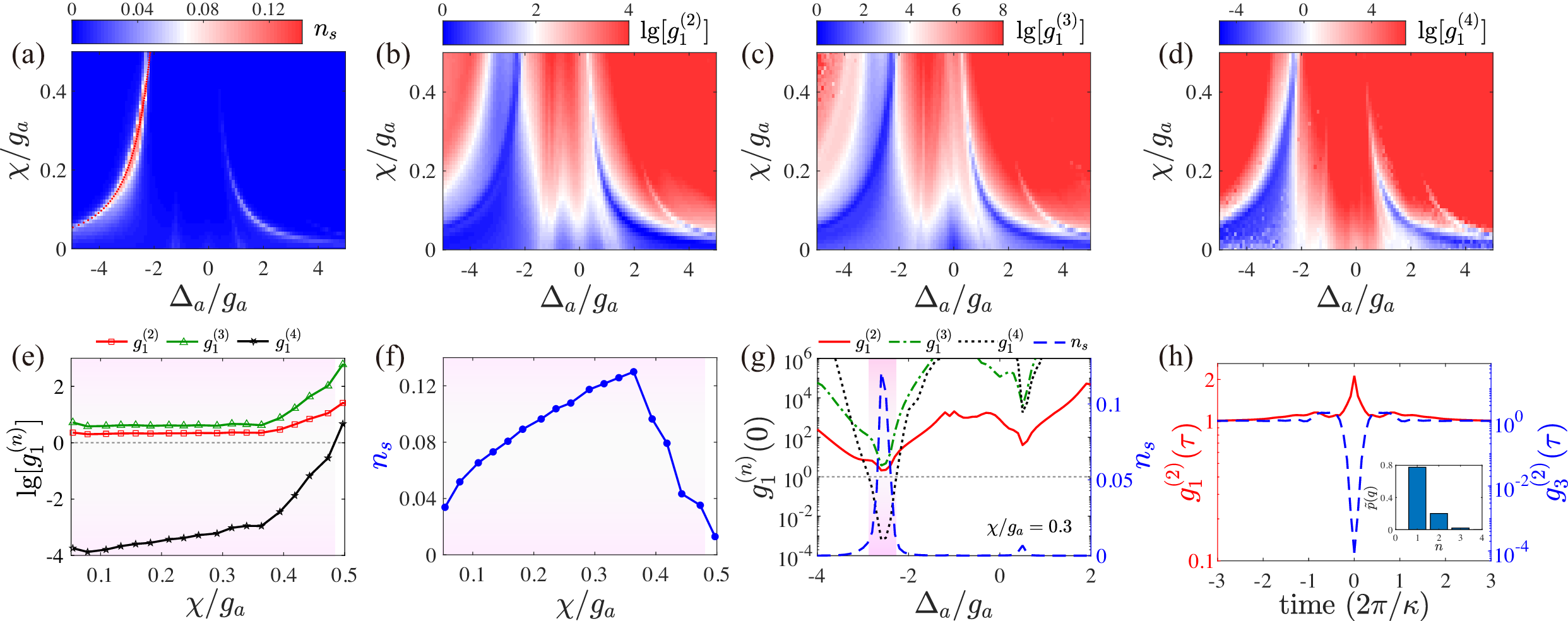}
\caption{Phase-space mapping of steady-state photon correlations and emission statistics in the $\chi$-$\Delta_a$ parameter plane for $\phi=2\pi/3$. 
(a-d) Steady-state photon number $n_s$ and equal-time second-, third-, and fourth-order correlation functions lg$[g_1^{(2)}(0)]$, lg$[g_1^{(3)}(0)]$, lg$[g_1^{(4)}(0)]$ as functions of the cavity detuning $\Delta_a$ and interaction strength $\chi$, with $\delta=(\chi/g_a)\Delta_a$. The red curve in (a) indicates the analytical three-photon resonance condition derived from the energy spectrum, showing excellent agreement with the numerical results.
(e,f) Extracted values of lg$[g_1^{(n)}(0)]$ and photon number $n_{s}$ along the resonance branch as functions of $\chi$, revealing the emergence of the three-photon emission regime. 
(g) Correlation functions $g_1^{(n)}(0)$ and photon number $n_s$ as functions of $\Delta_a$ at $\chi/g_a=0.3$, highlighting a regime of strong three-photon emission with a sizable photon population.
(h) Time-dependent correlations $g_1^{(2)}(\tau)$ (red) and $g_3^{(2)}(\tau)$ (blue) evaluated at the three-photon resonance point $\Delta_a/g_a=-2.62$. 
The inset shows $\tilde{p}(q)$ at the same point, demonstrating dominant three-photon occupation with strong suppression of single- and two-photon processes.}
\label{V_Deltac_2/3pi}
\end{figure*}

To further elucidate the underlying mechanism, we extract the photon statistics along the single-photon (red) and two-photon (white) resonance branches, as shown in Figs.~\ref{V_Deltac}(d,e). Along the single-photon resonance [Fig.~\ref{V_Deltac}(d)], both $g_1^{(2)}(0)$ and $g_1^{(3)}(0)$ decrease monotonically with increasing $\Delta_a$, while $n_s$ exhibits a nonmonotonic behavior. For $\Delta_a/g_a>0.9$, the system enters a well-defined single-photon regime with progressively enhanced purity. This reflects the increasing spectral separation between the $N=1$ manifold and higher-excitation manifolds, which suppresses multiphoton leakage. The nonmonotonic behavior of $n_s$ arises from a trade-off between excitation efficiency and spectral selectivity.

In contrast, along the two-photon resonance [Fig.~\ref{V_Deltac}(e)], $g_1^{(2)}(0)$ and $g_1^{(3)}(0)$ exhibit a nonmonotonic dependence on $\Delta_a$, signaling a continuous crossover from single-photon to two-photon emission. For $0.85<\Delta_a/g_a<2.05$, the residual coupling to the $N=1$ manifold remains significant, and the emission retains a single-photon character. Once $\Delta_a/g_a>2.05$, the $N=2$ channel becomes dominant and the system transitions into the two-photon regime. This crossover originates from the gradual enhancement of spectral isolation: increasing $\Delta_a$ pushes the $N=1$ manifold off resonance while isolating the $N=2$ manifold, thereby transferring spectral weight to the two-photon channel.

As shown in Figs.~\ref{V_Deltac}(f) and \ref{V_Deltac}(g), the optimal two-photon bundle emission is achieved at $\Delta_a/g_a=2.92$ with $\chi/g_a=1$, where $g_1^{(2)}(0)=1.21$, $g_1^{(3)}(0)=2\times10^{-4}$, and $n_s=0.1$. Compared to the noninteracting case, this corresponds to an enhancement of nearly three orders of magnitude in two-photon purity while maintaining a sizable photon population. This optimal point reflects a balance between spectral isolation and excitation efficiency.

To confirm the bundle nature of the emission, we analyze the time-dependent correlations $g_1^{(2)}(\tau)$ and $g_2^{(2)}(\tau)$ at this operating point [Fig.~\ref{V_Deltac}(h)]. The conditions $g_1^{(2)}(0)>g_1^{(2)}(\tau)$ and $g_2^{(2)}(0)<g_2^{(2)}(\tau)$ are simultaneously satisfied, demonstrating bunching within photon pairs and antibunching between successive pairs. The photon-number distribution further supports this interpretation, with $P(1)=0.89$ and $P(2)=0.11$, indicating that emission is dominated by two-photon events with strongly suppressed higher-order contributions. These results provide unambiguous evidence for genuine two-photon bundle emission.

Physically, this optimal regime originates from interaction-enhanced spectral anharmonicity: the cavity-mediated interaction $\chi$ isolates the $N=2$ manifold from neighboring excitation manifolds, while interference suppresses competing pathways. As a result, excitation is funneled into the two-photon channel, enabling simultaneous enhancement of purity and stability. This demonstrates that the interplay between interference and interaction enables not only the selection of excitation manifolds, but also their optimal stabilization, providing a robust route toward high-quality photon-pair sources. Importantly, this mechanism does not rely on intrinsically strong nonlinearities, but instead exploits interaction-engineered spectral control for programmable quantum light generation.

\subsection{Interaction-enhanced three-photon bundle emission}

We finally turn to the regime of three-photon emission for $\phi=2\pi/3$, where destructive interference selectively isolates the $N=3$ manifold. In this configuration, the optimal three-photon response is realized by tuning the system along the condition $\delta=(\chi/g_a)\Delta_a$, under which the $N=3$ manifold becomes resonant while the $N=1$ and $N=2$ channels remain effectively suppressed.

Figures~\ref{V_Deltac_2/3pi}(a-d) show the steady-state photon number $n_s$ and the correlation functions $g_1^{(2)}(0)$, $g_1^{(3)}(0)$, and $g_1^{(4)}(0)$ (in a  in logarithmic scale) as functions of $\Delta_a$ and $\chi$. The analytical three-photon resonance condition (red curve) agrees well with the numerical results, confirming the manifold-resolved picture. Along this resonance branch, the photon statistics exhibit a clear transition into a three-photon-dominated regime.

To quantify this behavior, we extract the photon statistics along the resonance curve, as shown in Figs.~\ref{V_Deltac_2/3pi}(e) and \ref{V_Deltac_2/3pi}(f). As the interaction strength $\chi$ increases, the system enters a stable regime where $g_1^{(2)}(0)$ and $g_1^{(3)}(0)$ remain nearly constant at $\sim3$ and $\sim4$, respectively, while $g_1^{(4)}(0)$ increases gradually. This plateau indicates a robust three-photon-dominated regime, where the $N=3$ manifold is well isolated and lower-order excitation channels remain effectively suppressed.

Meanwhile, the photon number $n_s$ exhibits a nonmonotonic dependence on $\chi$, reflecting a competition between excitation efficiency and spectral selectivity. Together, these results identify an optimal interaction window in which three-photon correlations are maximized while higher-order processes remain strongly suppressed. 

We therefore select $\chi/g_a=0.3$ as an optimal operating point to further characterize the three-photon bundle emission. As shown in Fig.~5(g), at $\Delta_a/g_a=-2.6$ the system exhibits strong three-photon emission with $g_1^{(2)}(0)=2.12$, $g_1^{(3)}(0)=3.74$, $g_1^{(4)}(0)=6.41\times10^{-4}$, and $n_s=0.12$. Compared to the noninteracting case, this corresponds to an improvement of more than two orders of magnitude in three-photon purity while maintaining a relatively large photon population.

The bundle nature of the emission is further confirmed by the time-dependent correlations shown in Fig.~5(h). The conditions $g_1^{(2)}(0)>g_1^{(2)}(\tau)$ and $g_3^{(2)}(0)<g_3^{(2)}(\tau)$ are simultaneously satisfied, demonstrating strong bunching within three-photon groups and antibunching between successive groups. The photon-number distribution further supports this interpretation, showing dominant three-photon occupation with strong suppression of single- and two-photon components.

Physically, this regime originates from the cooperative action of interference and interaction. The phase $\phi=2\pi/3$ suppresses the $N=1$ and $N=2$ excitation pathways via destructive interference, while the cavity-mediated interaction $\chi$ enhances spectral anharmonicity and isolates the $N=3$ manifold. As a result, excitation is funneled into the three-photon channel, enabling simultaneous enhancement of emission purity and efficiency.

Together with the single- and two-photon regimes discussed above, these results establish a unified mechanism for programmable few-photon emission: phase-controlled interference selects the target excitation manifold, while interaction-induced spectral anharmonicity stabilizes and optimizes the corresponding emission process.Importantly, this mechanism enables high-purity multiphoton emission without relying on intrinsically strong nonlinearities, highlighting the effectiveness of interaction-engineered spectral control.

\section{Conclusion}
In summary, we have developed a unified interference-interaction mechanism for programmable few-photon emission in a cavity-QED system. By combining phase-controlled interference with the tunable cavity-mediated SEI, we demonstrate selective addressing and stabilization of excitation manifolds, enabling deterministic control of photon-emission channels. Within this framework, constructive interference enhances the spectral anharmonicity of low-excitation manifolds, giving rise to high-purity single- and two-photon emission, whereas destructive interference suppresses lower-order pathways and activates higher-order processes, enabling three-photon bundle emission. The cavity-mediated interaction further amplifies spectral separation between manifolds, leading to simultaneous enhancement of emission purity and photon population. As a result, both two- and three-photon emissions exhibit orders-of-magnitude improvements in purity without requiring intrinsically strong optical nonlinearities. 

These results establish a general paradigm in which optical nonlinearity is not fixed by material properties but can be programmably engineered through the interplay of interference and many-body interactions. The required parameter regime is readily accessible in state-of-the-art platforms, including single-atom cavity QED, Rydberg cavity systems, and alkaline-earth-atom implementations~\cite{schuck2016quantum, jhuria2024programmable}. More broadly, our work connects to recent advances in programmable quantum systems and synthetic quantum matter~\cite{ebadi2021quantum, chen2022synthetic, yang2024programmable}, demonstrating that quantum optical responses can be coherently controlled via phase and effective interactions. Together with recent progress in integrated quantum photonics~\cite{elshaari2020hybrid,wang2020integrated,luo2023recent}, these results provide a scalable route toward phase-engineered quantum light sources. The proposed mechanism naturally extends to multi-emitter arrays, multimode cavities, and networked architectures~\cite{stokowski2023integrated, zhu2024dynamically}, opening new opportunities for controllable many-body quantum optics and quantum information processing.

\begin{acknowledgments}
This work was supported by the National Natural Science Foundation of China (Grant No.12374365, Grant No. 12274473, and Grant No. 12135018), Guangdong Provincial Quantum Science Strategic Initiative (Grants No. GDZX2505001), and Guangdong University of Technology SPOE Seed Foundation (SF2024111504).  
\end{acknowledgments}

\begin{center}
\textbf{\uppercase{Data availability}}
\end{center}

\noindent
The data that support the findings of this article are not publicly available. 
The data are available from the authors on reasonable request.


\appendix

\section{Energy spectrum for cavity-coupled atomic array\label{sm_model}}

In this appendix, we derive the analytical zero-energy resonant condition for cavity-coupled atomic array in different excitation manifolds. These conditions define the manifold-resolved resonance curves that govern the photon-emission processes discussed in the main text. For ignoring the weak pump field ($\Omega=0$), the zero-energy resonances are obtained by projecting the Hamiltonian (\ref{Hami2}) onto fixed excitation-number manifolds and solving: $\det(\mathcal{M}_N)=0$. Here $\mathcal{M}_N$ is the Hamiltonian matrix in the $N$-excitation subspace. 

\subsection{Single-excitation manifold with $\phi=0$}
We first analyze the $N=1$ manifold for fixing $\phi=0$ and $\delta=\Delta_a/2$. Then the zero-energy resonant condition can be obtained analytically due to the underlying $S_3$ symmetry, corresponding to the Hamiltonian in the basis $\Psi = \{|gge,0\rangle, |geg,0\rangle, |egg,0\rangle, |ggg,1\rangle\}^T$ \begin{equation}
\mathcal{M}_1=
\begin{pmatrix}
\frac{\Delta_a}{2}+\chi & \chi & \chi & g_a\\
\chi & \frac{\Delta_a}{2}+\chi & \chi & g_a\\
\chi & \chi & \frac{\Delta_a}{2}+\chi & g_a\\
g_a & g_a & g_a & \Delta_a
\end{pmatrix},
\end{equation}
The $S_3$ permutation symmetry allows the atomic subspace to be decomposed into two dark states and one fully symmetric bright state. The collective symmetric subspace $\{|W,0\rangle, |ggg,1\rangle\}$ is governed by the reduced Hamiltonian
\begin{equation}
\mathcal{M}_{\mathrm{coll}}=
\begin{pmatrix}
\frac{\Delta_a}{2}+3\chi & \sqrt{3}g_a\\
\sqrt{3}g_a & \Delta_a
\end{pmatrix},
\end{equation}
where $|W\rangle=(|gge\rangle+|geg\rangle+|egg\rangle)/\sqrt{3}$.

Imposing the zero-energy condition $\det(\mathcal{M}_{\mathrm{coll}})=0$ yields 
\begin{equation}
\Delta_a=-3\chi\pm\sqrt{9\chi^2+6g_a^2},
\end{equation}
For $\phi=0$ and $\delta=-\Delta_a/2$, the same procedure leads to 
\begin{equation}
\Delta_a = 3\chi \pm \sqrt{9\chi^2 - 6g_a^2}.
\end{equation}
 These results show that the cavity-mediated spin-exchange interaction $\chi$ shifts and reshapes the vacuum Rabi splitting, providing a tunable spectral structure.

\subsection{Double-excitation manifold  with $\phi=0$}
We now turn to the $N=2$ manifold. For $\phi=0$ and $\delta=\Delta_a/2$, the Hilbert space is spanned by seven basis states, excluding the inaccessible $|-1,e,e,e\rangle$. The Hamiltonian takes the form
\begin{widetext}
\begin{equation}
\mathcal{M}_{2}=
\begin{pmatrix}
\frac{\Delta_a}{2}+\chi & \chi & \chi & 0 & g_a & g_a & \sqrt{2}g_a\\
\chi & \frac{\Delta_a}{2}+\chi & \chi & g_a & 0 & g_a & \sqrt{2}g_a\\
\chi & \chi & \frac{\Delta_a}{2}+\chi & g_a & g_a & 0 & \sqrt{2}g_a\\
0 & g_a & g_a & -\Delta_a+2\chi & \chi & \chi & 0\\
g_a & 0 & g_a & \chi & -\Delta_a+2\chi & \chi & 0\\
g_a & g_a & 0 & \chi & \chi & -\Delta_a+2\chi & 0\\
\sqrt{2}g_a & \sqrt{2}g_a & \sqrt{2}g_a & 0 & 0 & 0 & 2\Delta_a
\end{pmatrix},
\end{equation}
\end{widetext}
The Hamiltonian retains the $S_3$ symmetry and can be reduced to a collective symmetric subspace. In the basis $\{|W_1\rangle, |W_2\rangle, |G_2\rangle\}$, the reduced Hamiltonian reads
\begin{equation}
\mathcal{M}_{\mathrm{coll}}^{(2)} =
\begin{pmatrix}
\frac{3\Delta_a}{2}+3\chi & 2g_a & \sqrt{6}\,g_a \\
2g_a & \Delta_a+4\chi & 0 \\
\sqrt{6}\,g_a & 0 & 2\Delta_a
\end{pmatrix},
\label{Mcoll_n2}
\end{equation}
Explicitly, the symmetric states are defined as
\begin{subequations}
\begin{align}
|W_1\rangle &= \frac{1}{\sqrt{3}}(|e,g,g,1\rangle + |g,e,g,1\rangle + |g,g,e,1\rangle), \\
|W_2\rangle &= \frac{1}{\sqrt{3}}(|e,e,g,0\rangle + |e,g,e,0\rangle + |g,e,e,0\rangle),
\end{align}
\end{subequations}
where $|W_1\rangle$ and $|W_2\rangle$ represent the symmetric states with one and two atomic excitations, respectively, while $|G_2\rangle$ denotes the pure two-photon cavity state.

Then the zero-energy condition $\det(\mathcal{M}_{\mathrm{coll}}^{(2)})=0$ leads to
\begin{equation}
3\Delta_a^3
+18\chi \Delta_a^2
+\left(24\chi^2-14g_a^2\right)\Delta_a
-24g_a^2\chi
=0,
\label{zero_n2}
\end{equation}
For $\delta=-\Delta_a/2$, the corresponding condition becomes
\begin{equation}
\Delta_a^3+2\chi\Delta_a^2+(2g_a^2-24\chi^2)\Delta_a+24g_a^2\chi=0.
\end{equation}

\subsection{Triple-excitation manifold with $\phi=2\pi/3$}
For $\phi=2\pi/3$, destructive interference suppresses the lower-order excitation pathways and enables resonant higher-order processes. The system is spanned by the basis 
\[
\begin{aligned}
\Psi = \{ & |2,g,g,e\rangle, |2,g,e,g\rangle, |2,e,g,g\rangle, \\
& |1,e,e,g\rangle, |1,e,g,e\rangle, |1,g,e,e\rangle, \\
& |0,e,e,e\rangle, |3,g,g,g\rangle \}^T,
\end{aligned}
\]

By introducing $\omega=e^{i2\pi/3}$ with $\omega^*=\omega^{-1}=e^{-i2\pi/3}$, the Hamiltonian in $n=3$ excitation manifold can be written as
\begin{widetext}
\begin{equation}
\mathcal{M}_3=
\begin{pmatrix}
E_1 & \chi & \chi & 0 & \sqrt{2}g_a\omega & \sqrt{2}g_a & 0 & \sqrt{3}g_a\omega\\
\chi & E_1 & \chi & \sqrt{2}g_a\omega & 0 & \sqrt{2}g_a\omega^* & 0 & \sqrt{3}g_a\\
\chi & \chi & E_1 & \sqrt{2}g_a & \sqrt{2}g_a\omega^* & 0 & 0 & \sqrt{3}g_a\omega^*\\
0 & \sqrt{2}g_a\omega^* & \sqrt{2}g_a & E_2 & \chi & \chi & g_a\omega^* & 0\\
\sqrt{2}g_a\omega^* & 0 & \sqrt{2}g_a\omega & \chi & E_2 & \chi & g_a & 0\\
\sqrt{2}g_a & \sqrt{2}g_a\omega & 0 & \chi & \chi & E_2 & g_a\omega & 0\\
0 & 0 & 0 & g_a\omega & g_a & g_a\omega^* & E_3 & 0\\
\sqrt{3}g_a\omega^* & \sqrt{3}g_a & \sqrt{3}g_a\omega & 0 & 0 & 0 & 0 & E_4
\end{pmatrix},
\label{M3_phi23}
\end{equation}
\end{widetext}
where
\begin{align}
E_1&=\frac{5}{2}\Delta_a+\chi,\nonumber \\
E_2&=2\Delta_a+2\chi,\nonumber \\
E_3&=\frac{3}{2}\Delta_a+3\chi,\nonumber \\
E_4&=3\Delta_a,
\end{align}
for $\delta=\Delta_a/2$. The zero-energy condition $\det(\mathcal{M}_3)=0$ can be factorized as
\begin{align}
\det(\mathcal{M}_3)\propto&
\left(5\Delta_a^2+10\chi\Delta_a-2g_a^2\right)\nonumber\\
&\times
\left(10\Delta_a^2+17\chi\Delta_a+6\chi^2-4g_a^2\right)
\nonumber\\
&\times
\Big(
10\Delta_a^4+25\chi\Delta_a^3+10\chi^2\Delta_a^2
-38g_a^2\Delta_a^2
\nonumber\\
&\qquad
-62\chi g_a^2\Delta_a
-12\chi^2 g_a^2
+12g_a^4
\Big),
\label{detM3_factorized}
\end{align}

Therefore, the zero-energy conditions in the $n=3$ manifold consist of three distinct branches:
\begin{align}
5\Delta_a^2+10\chi\Delta_a-2g_a^2 &= 0,
\label{n3_branch1}
\\
10\Delta_a^2+17\chi\Delta_a+6\chi^2-4g_a^2 &= 0,
\label{n3_branch2}
\\
10\Delta_a^4+25\chi\Delta_a^3+10\chi^2\Delta_a^2
-38g_a^2\Delta_a^2
-62\chi g_a^2\Delta_a
\nonumber\\
\qquad\qquad
-12\chi^2 g_a^2
+12g_a^4 &= 0.
\label{n3_branch3}
\end{align}
These equations determine the zero-energy resonant branches of $n=3$ excitation manifold.  The two quadratic branches originate from interference-selected subspaces, while the quartic branch describes the collective bright sector. Their coexistence reflects the richer dressed-state structure in $n=3$ manifold and underlies the emergence of phase-selected three-photon resonance.
%

\end{document}